\documentclass[12pt]{article} 

\usepackage{OL}
\usepackage{hyperref}
\usepackage{amsmath}
\usepackage{epsfig}
\graphicspath{{figs/}}

\begin{document}

\title{Introduction to microstructured optical fibers}

\author{M. Yan}

\address{myan@ieee.org \\ Network Technology Research Centre, Nanyang Technological University \\ 50 Nanyang Drive, Singapore, 637553}

\begin{center}
{\bf{Abstract}}
\end{center}
This article recounts definition, classification, history, and
applications of microstructured optical fibers.

\section{Definition and Varieties}

Since the first microstructured optical fiber (MOF) demonstrated
by Knight \emph{et al.} \cite{Knight:PCF96}, research on this
class of optical waveguides has thrived. To date, lots of fibers
with air-hole inclusions were reported. In fact, to meet different
application purposes, the ``holey'' fibers experience constant
alterations and they evolve divergently in many directions. Some
of fibers deviate much in structure from that in the very first
publication. To address these fibers, several names have been
coined---they include \emph{photonic crystal fiber} (PCF),
\emph{holey fiber} (HF), \emph{microstructured optical fiber}
(MOF), and \emph{photonic bandgap fiber} (PBGF). ``PCF'', ``HF'',
and ``MOF'' are defined from structure point of view, whereas
``PBGF'' are defined from optical-property point of view. Among
the names, MOF is the most general one. In fact, MOF can be used
to address all fibers which have their feature size at micrometer
(or submicrometer) scale. The rest names can be considered as
subsets of MOF, and they can't address rigorously all MOFs
reported in general. For example, ``PCF'' is not appropriate for
fibers which lack of periodic cladding; ``HF'' is not appropriate
for those fibers made of all solid materials; ``PBGF'' is
certainly only applicable for those guiding light by cladding's
photonic bandgap (PBG).

\begin{figure}
\centering
\includegraphics[clip=true, width=6.4cm]{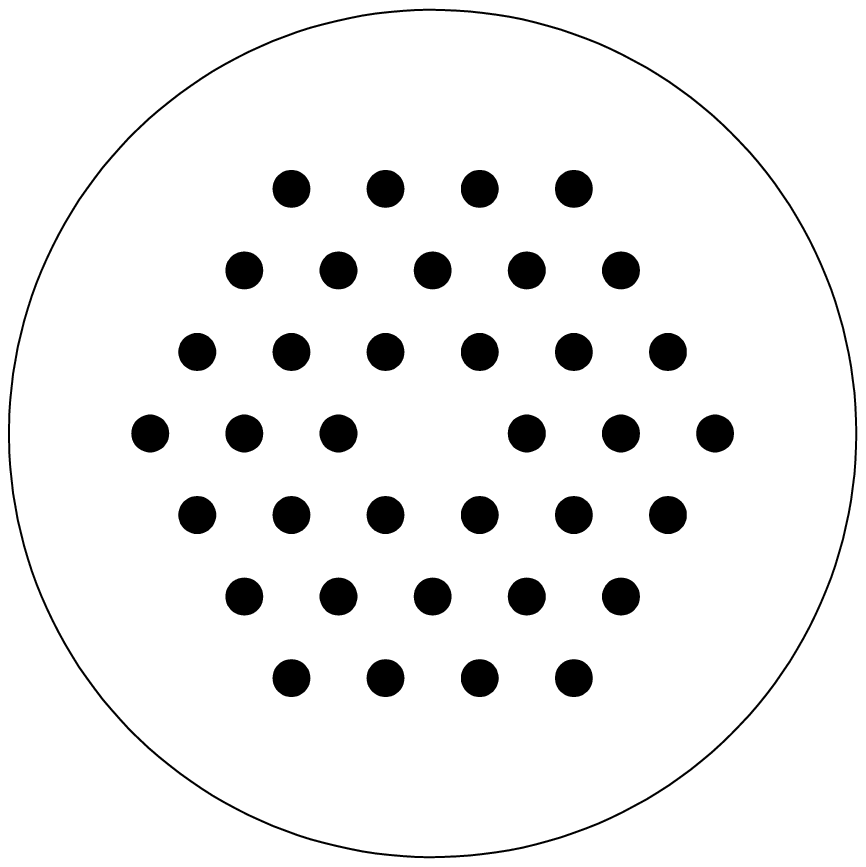}
\includegraphics[clip=true, width=6.4cm]{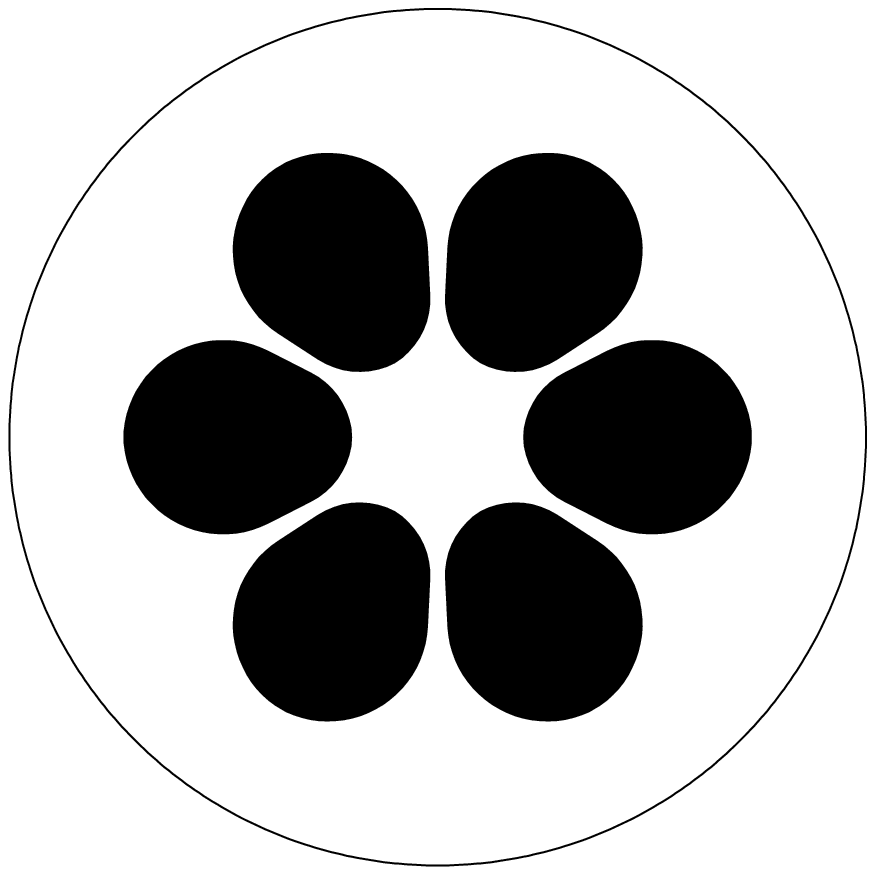}
\put(-282, -10){\textsf{(a)}}
\put(-98, -10){\textsf{(b)}}    \\
\includegraphics[clip=true, width=6.4cm]{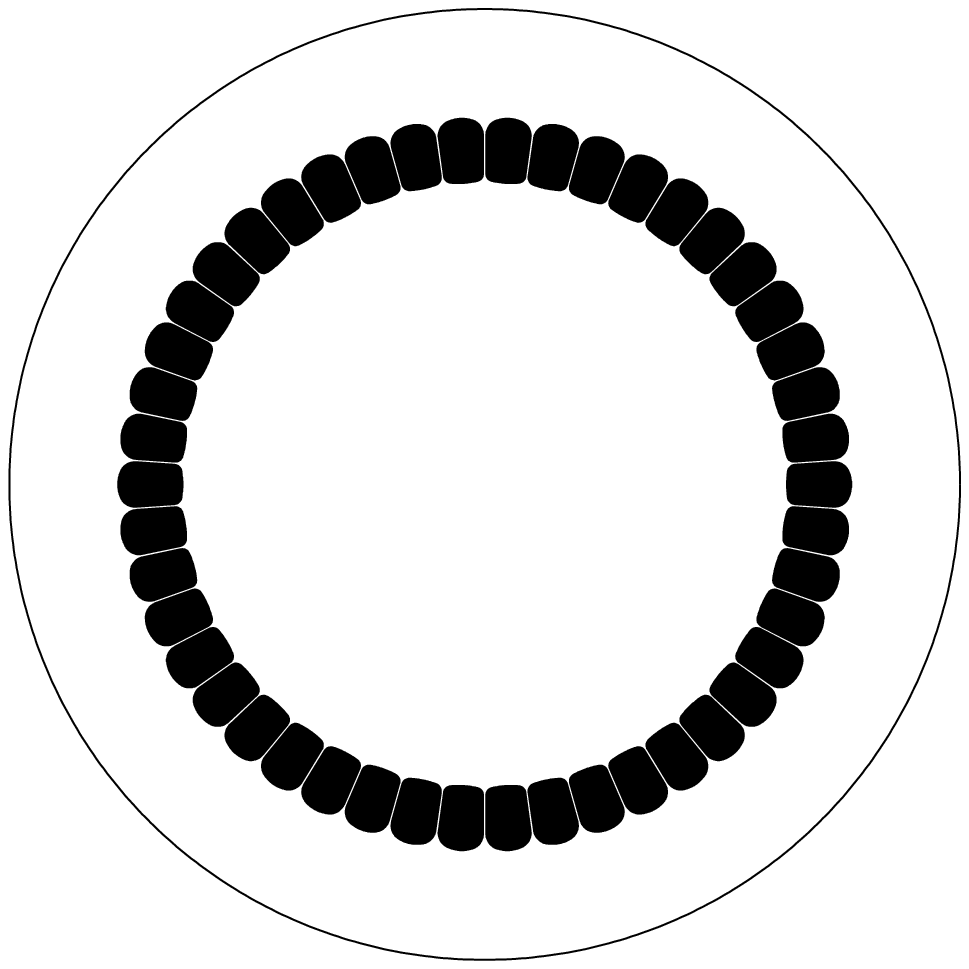}
\includegraphics[clip=true, width=6.4cm]{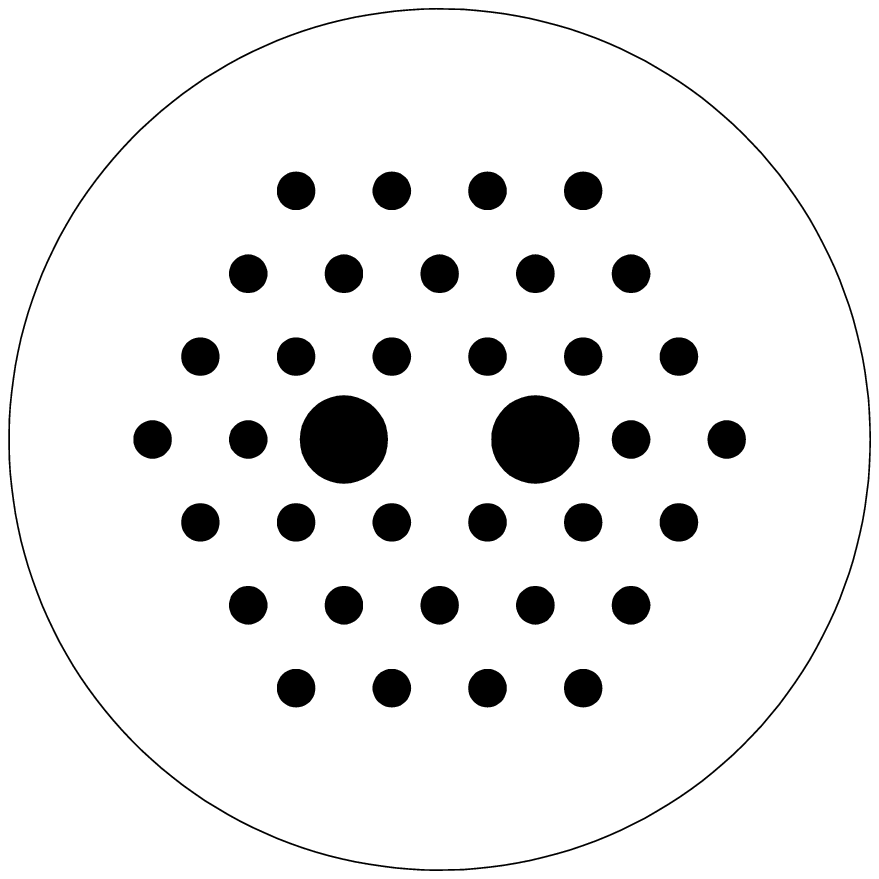}
\put(-282, -10){\textsf{(c)}}
\put(-98, -10){\textsf{(d)}}    \\
\includegraphics[clip=true, width=6.4cm]{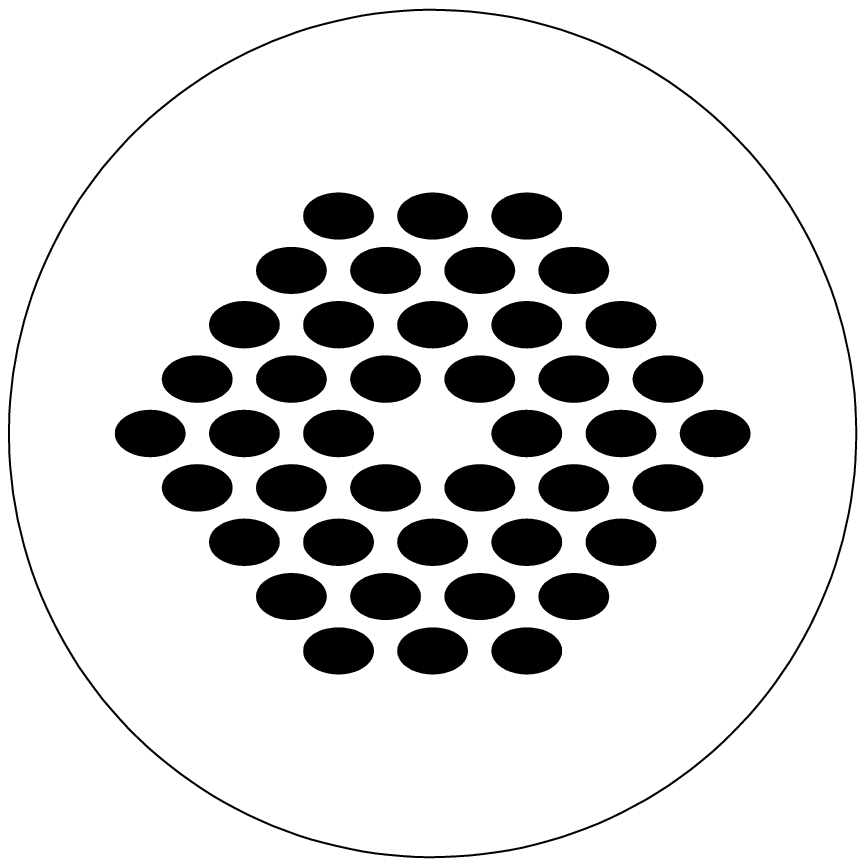}
\includegraphics[clip=true, width=6.4cm]{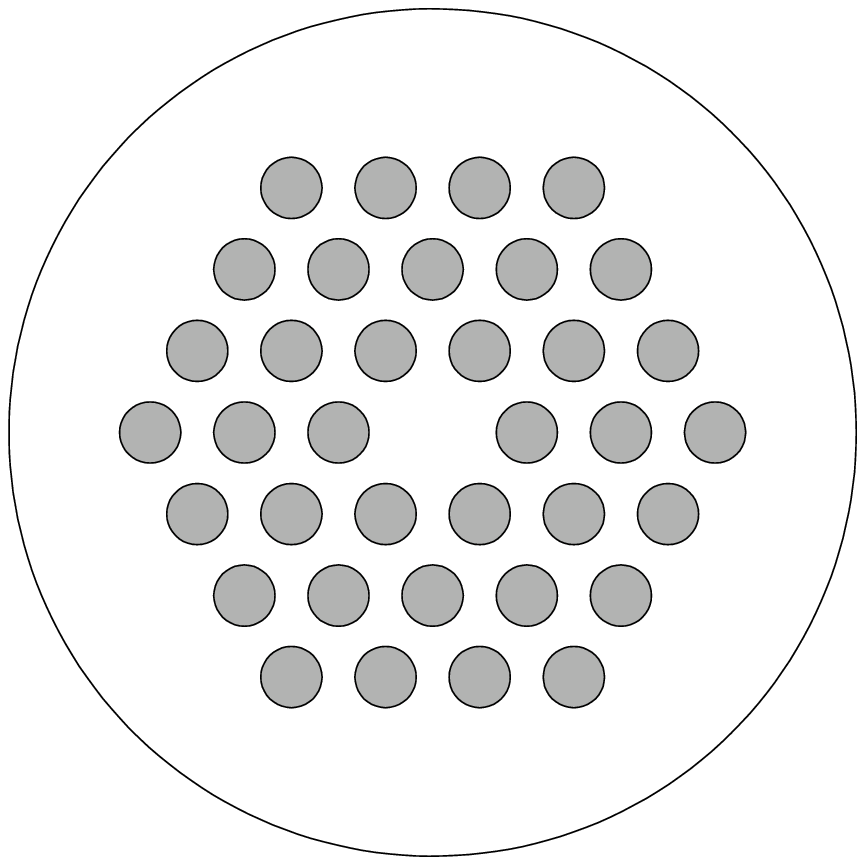}
\put(-282, -10){\textsf{(e)}}
\put(-98, -10){\textsf{(f)}}    \\
\caption{(a) Index-guiding PCF \cite{Knight:PCF96}. (b) MOF with
six holes \cite{Eggleton:MOFDevices}. (c) So-called air-clad
fiber, which has very large numerical aperture
\cite{Crystalfiber}. (d) High-birefringent MOF
\cite{Kubota:SinglePolarization}. (e) MOF with elliptic cladding
air holes \cite{Nader:EllipticHole}. (f) Index-guiding PCF with
low-index rods in cladding \cite{Feng:SOHO}. Black is for air,
white is for silica. In panel (f), grey is for a solid material
whose index is smaller than that of the background.}
\label{fig:MOFIG}
\end{figure}

\begin{figure}
\centering
\includegraphics[clip=true, width=6.4cm]{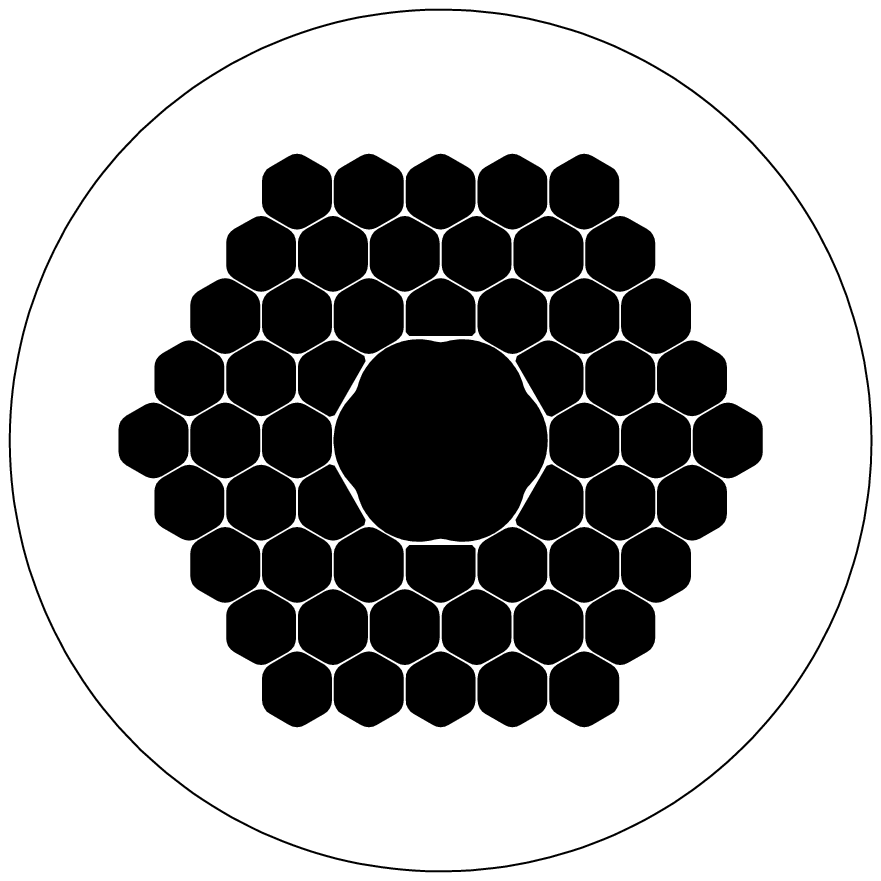}
\includegraphics[clip=true, width=6.4cm]{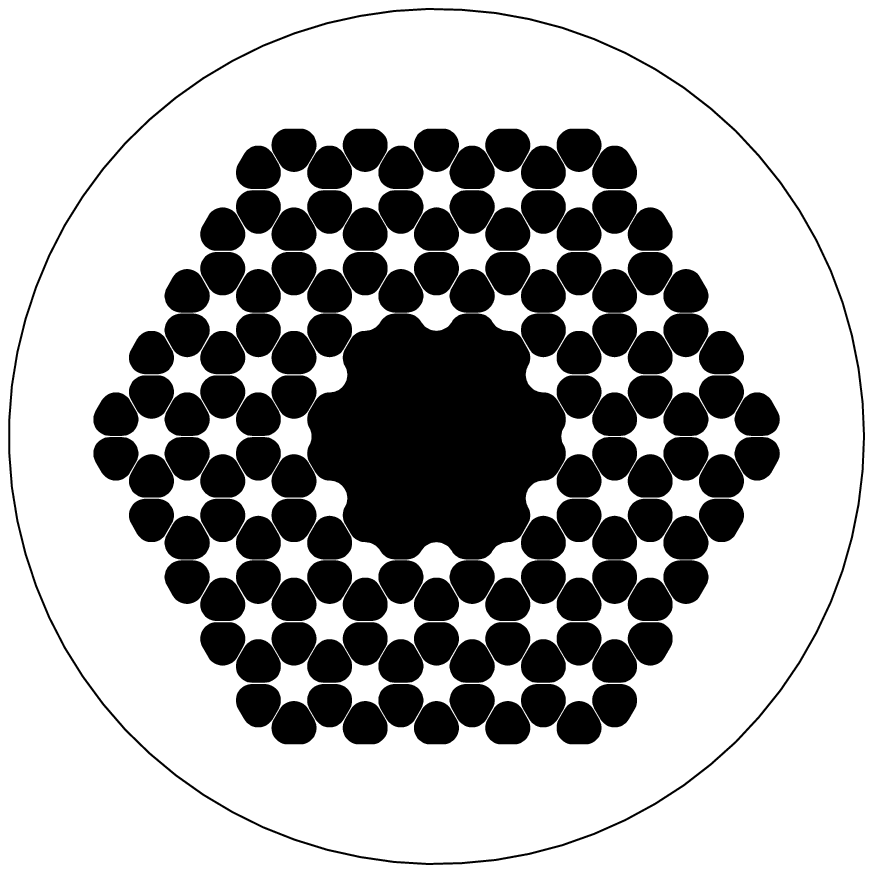}
\put(-282, -10){\textsf{(a)}}
\put(-98, -10){\textsf{(b)}}    \\
\includegraphics[clip=true, width=6.4cm, angle=90]{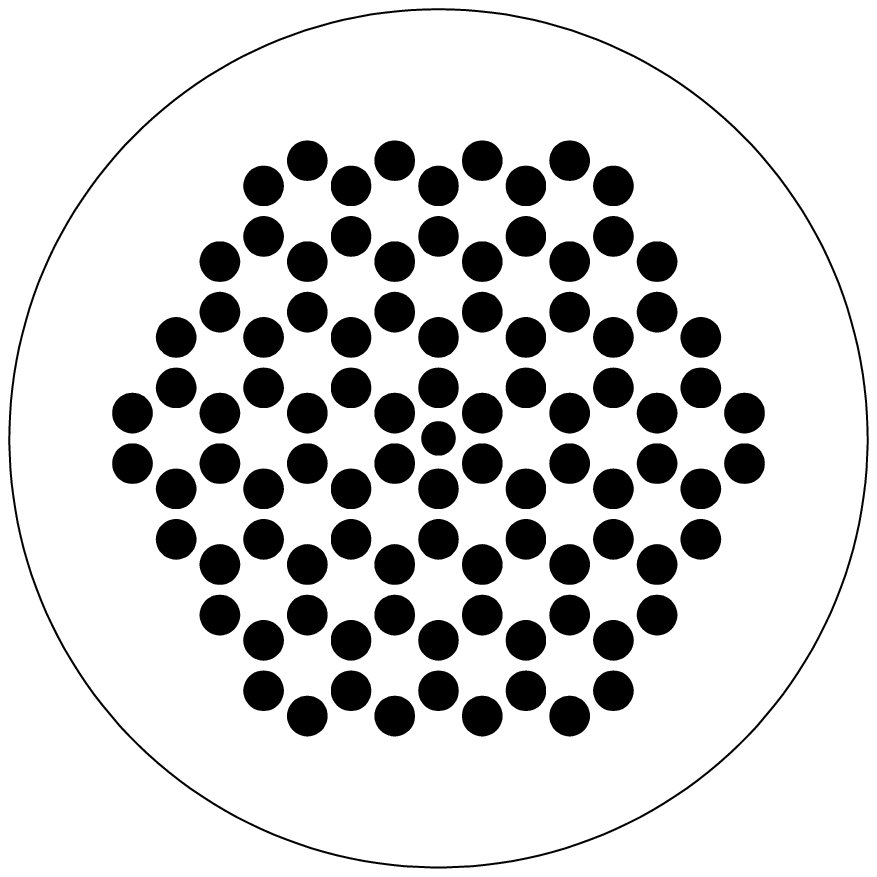}
\includegraphics[clip=true, width=6.4cm]{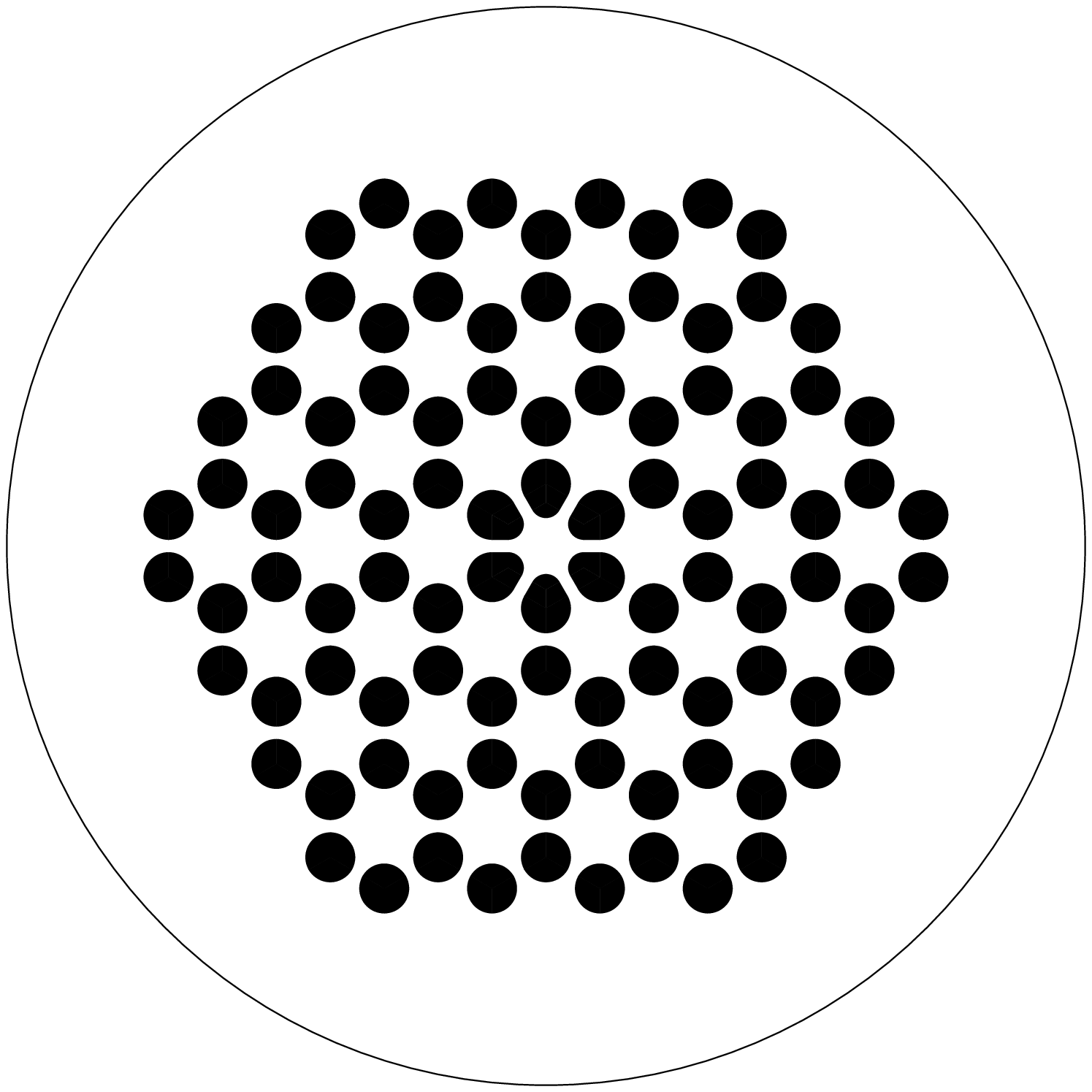}
\put(-282, -10){\textsf{(c)}}
\put(-98, -10){\textsf{(d)}}    \\
\includegraphics[clip=true, width=6.4cm]{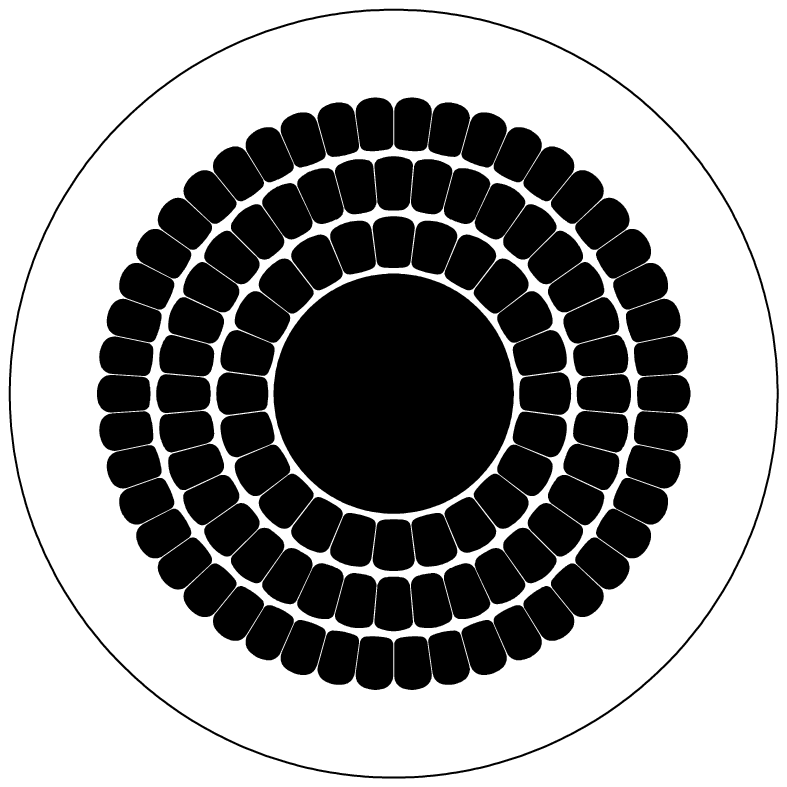}
\includegraphics[clip=true, width=6.4cm]{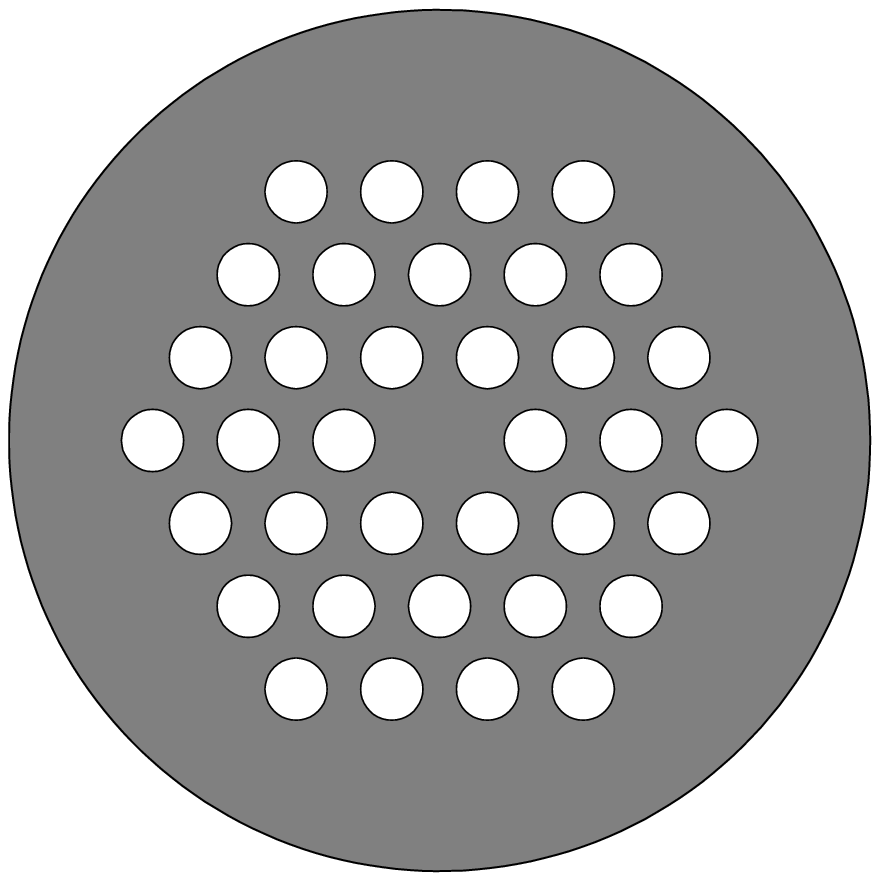}
\put(-282, -10){\textsf{(e)}}
\put(-98, -10){\textsf{(f)}}    \\
\caption{(a) PBGF whose cladding is of triangular lattice
\cite{Smith:TPCFCorning}. (b) PBGF whose cladding is of honeycomb
lattice \cite{Yan:HPCFAirCoreDesign}. (c) Traditional honeycomb
PBGF \cite{Broeng:HPCF}. (d) Improved honeycomb PBGF
\cite{Yan:HPCF}. (e) Air-silica Bragg fiber \cite{Vienne:BFOPEX}.
(f) All-solid PBGF \cite{Luan:AllSolidPBGF}. Black is for air,
white is for silica. In panel (f), white is for a solid material
whose index is larger than that of the background.}
\label{fig:MOFPBG}
\end{figure}

Figure \ref{fig:MOFIG} and \ref{fig:MOFPBG} show novel MOFs made
or proposed during recent years. While PBG fibers (except the
air-silica Bragg fiber) shown in Fig. \ref{fig:MOFPBG} are
observed to adhere to the term ``photonic crystal (PC)''
faithfully, index-guiding fibers sometimes can be nothing to do
with photonic crystal (for example, the air-clad fiber). We will
come across most of the fibers displayed in these two figures
later in Section \ref{Sec:Review} of this article.

Index-guiding MOFs \emph{not} presented in Fig. \ref{fig:MOFIG}
include hole-assisted MOFs \cite{Hasegawa:HoleAssisted,Yan:HAMRF},
double-clad index-guiding MOF \cite{Crystalfiber}, and
``graded-index'' MOF \cite{Eijkelenborg:GRINMOF}. PBG-guiding MOFs
\emph{not} presented in Fig. \ref{fig:MOFPBG} include hollow-core
Bragg fiber \cite{Temelkuran:BraggFiberNature} and solid-core
Bragg fiber \cite{Katagiri:SilicaCorePBGF}. Heterostructured PCFs
\cite{Yan:HSPCF} are not shown in either of the figures.

Having mentioned all these MOF varieties, we will however limit
discussions on MOFs with periodic cladding. Hence in certain
places an MOF can be also referred to as PCF without ambiguity.

\subsection{Index-guiding \emph{v.s.} PBG-guiding}

In fact, the index-guiding fiber can be considered as a
PBG-guiding fiber in the sense that the propagation constant of a
core mode can not be supported by the cladding. The bandgap region
used in an index-guiding fiber is the largest one possessed by the
cladding, \emph{i.e.}, the region below the cladding's radiation
line $\beta=kn_{\mbox{\small clad}}$ in $\beta-k$ plot ($\beta$ is
the $x$-axis value, $\beta=\frac{2\pi}{\lambda}n_{\mbox{\small
eff}}$ and $k=\frac{2\pi}{\lambda}$, with $\lambda$ free-space
wavelength). In this region, the cladding can't support any
propagating mode, as the lightwave (wavelength at $\lambda$) in a
homogeneous material (refractive index at $n$) should have its
smallest propagation constant at $\beta=\frac{2\pi}{\lambda}n$.
And the smallest propagation constant happens when the light
propagates as a plane wave. Any other mode of propagation will
increase its propagation constant.

\begin{figure}
\centering
\includegraphics[width=10cm]{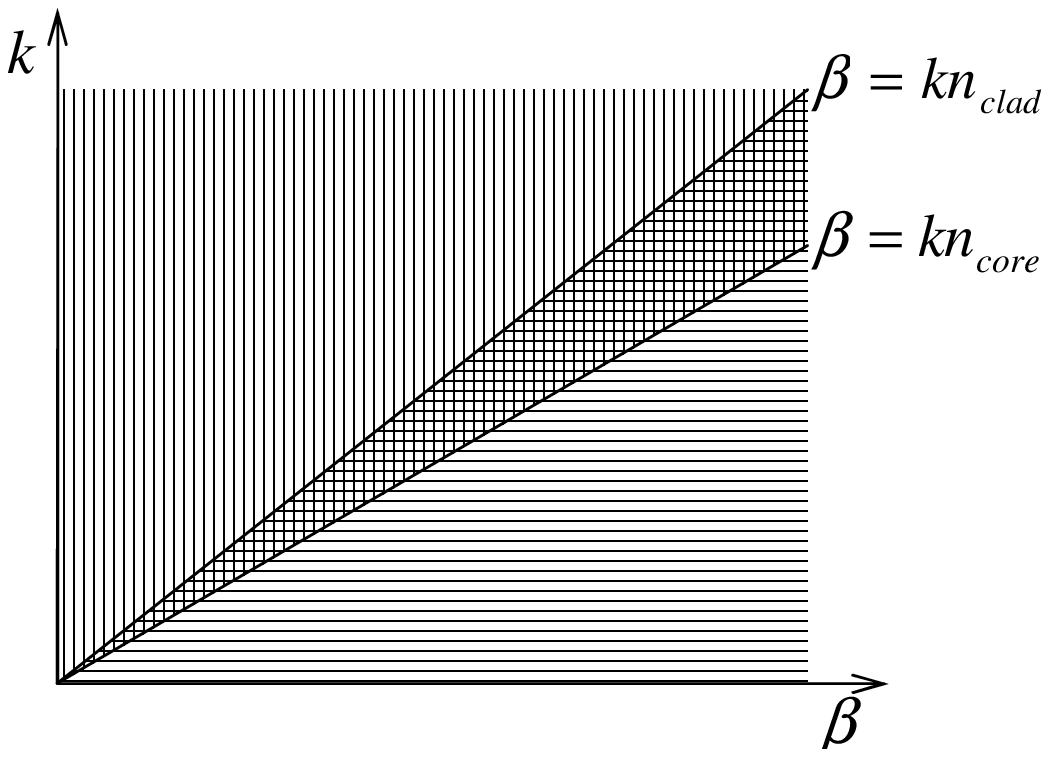}
\put(-160, -5){\textsf{\Large (a)}} \\
\includegraphics[width=10cm]{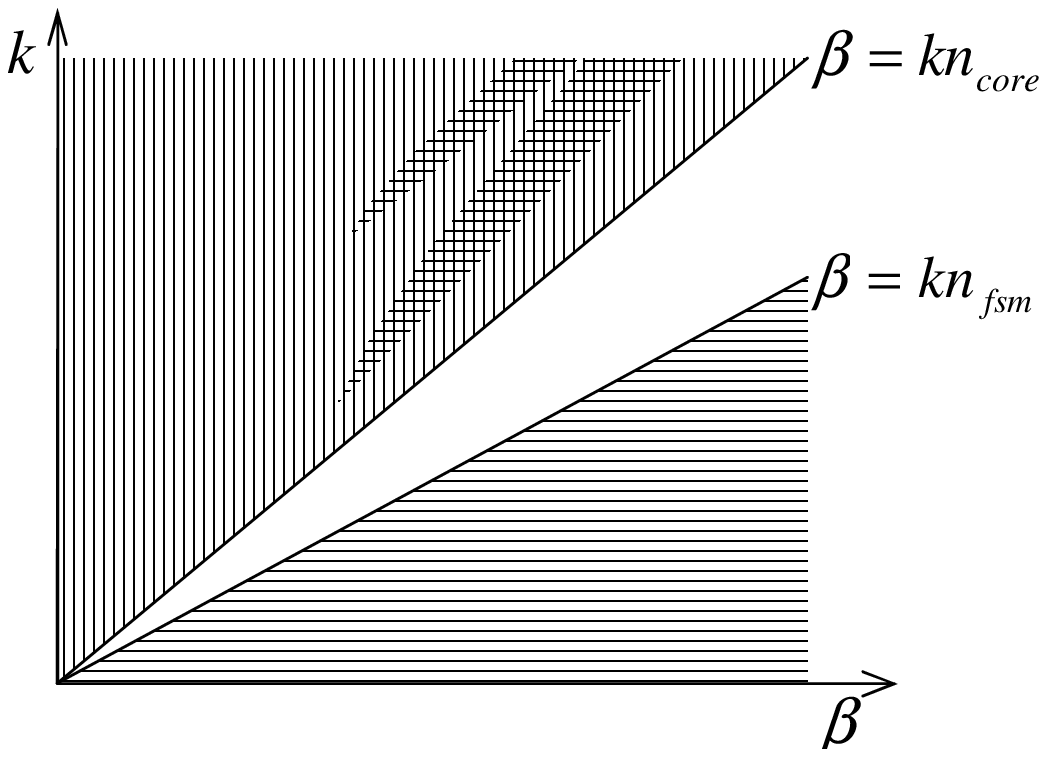}
\put(-160, -5){\textsf{\Large (b)}} \caption{Schematic dispersion
diagrams for: (a) A fiber with a low-index homogeneous cladding
and a high-index homogeneous core; (b) A fiber with photonic
crystal cladding and a homogeneous core. Regions filled with
horizontal lines are lightwave states forbidden in the cladding
material. Region filled with vertical lines are lightwave states
allowed in the core material. There common regions (grid-shaded
regions) are where confined states (allowed in core but not
allowed in cladding) are possible. $n_{\mbox{\small core}}$,
$n_{\mbox{\small clad}}$, $n_{\mbox{\small fsm}}$ are assumed to
be constant with respect to wavelength.}\label{fig:betak}
\end{figure}

Refer to Fig. \ref{fig:betak}, as an infinitely-extending rod of
high-index ($n_{\mbox{\small core}}$) is introduced in an
infinitely-extending low-index material ($n_{\mbox{\small
clad}}$), due to the fact that the high-index material has its
radiation line $\beta=kn_{\mbox{\small core}}$ positioned lower
than that for the low-index material $\beta=kn_{\mbox{\small
clad}}$, there exists a region in the $\beta-k$ plot (enclosed by
$\beta=k_n{\mbox{\small core}}$ and $\beta=kn_{\mbox{\small
clad}}$) in which light can propagating in the rod but not in the
background material. This is how a conventional step-index fiber
works.

Now as the homogeneous background material is replaced by a
photonic crystal, the cladding's radiation line becomes
$\beta=kn_{\mbox{\small fsm}}$, where $n_{\mbox{\small fsm}}$ is
effective index of the fundamental space-filling mode (FSM) of the
crystal structure. The main difference between a composite and a
homogeneous material is that, other than the region below its
radiation line, there are, possibly, some small regions above the
radiation line, in which light can't propagate in the crystal
[Fig. \ref{fig:betak}(b)]. Indeed, we can make use of these small
regions to confine light in a material whose index is smaller than
$n_{\mbox{\small fsm}}$. This is how a photonic bandgap fiber
works.

\subsection{PBG-guiding \emph{v.s.} Guidance by Antiresonant-reflection}

To differentiate these two terms, we have to recall the history of
the \emph{antiresonant-reflecting optical waveguide} (ARROW).

Multi-layered substrate was proposed by Ash to replace homogeneous
substrate in slab waveguide to guide surface wave in 1970
\cite{Ash:GratingWaveguides} \footnote{This paper is cited in
\cite{Fox:GratingGuide}, but is not traceable at this moment by
me.}. Such problem was then theoretically treated by A. J. Fox in
1974 \cite {Fox:GratingGuide}. Both papers call such waveguide as
grating waveguide. However, Fox did not show wave guidance in a
low-index material. Two years later, Yeh \emph{et al.}
theoretically demonstrated, by use of Floquet-Bloch theorem,
guidance in low-index material (air) for a slab waveguide with
layered cladding \cite{Yeh:Multilayered}. Such waveguide is named
by Yeh \emph{et al.} as Bragg waveguide. In this paper, very
importantly, they used a third material in core region (air), and
cladding is a periodic structure made of two solid materials. Half
a year later, Cho from Bell Laboratories experimentally confirmed
confined propagation in such waveguides \cite{Cho:BraggWG}. In
1978, Yeh \emph{et al.} put one step further from slab-type
waveguides, and proposed to use multi-layered cylinders to
propagate light in an air column \cite{Yeh:BraggFiber}. Such fiber
is named as Bragg fiber. Both slab-type Bragg waveguide and the
Bragg fiber were not investigated further until eight years later
in 1986 Duguay \emph{et al.} fabricated on silicon wafer a Bragg
waveguide \cite{Duguay:ARROW}. Their waveguide is however formed
by two material, \emph{i.e.}, core material is one of the cladding
materials. But Duguay \emph{et al.} rename the waveguide as
``antiresonant-reflecting optical waveguide'' (ARROW) and
attribute the guidance to the antiresonance of the high-index
layer (analogous to a Fabry-Perot resonator) placed adjacent to
the low-index core. To some extent, this paper revives the
research on such waveguides. A 2D ridge-type ARROW was also
fabricated by Freye \emph{et al.} in 1994 \cite{Freye:ARROW2D}.

Research on ARROW should really be considered as pioneer work on
photonic crystal waveguides (bandgap guidance). In fact, these
work can easily lead to the concept of photonic crystal. However,
as the authors did not generalize the \emph{layered dielectric
media} to other dimensions, their impact was limited.

\begin{figure}
\centering
\includegraphics[clip=true, width=8cm]{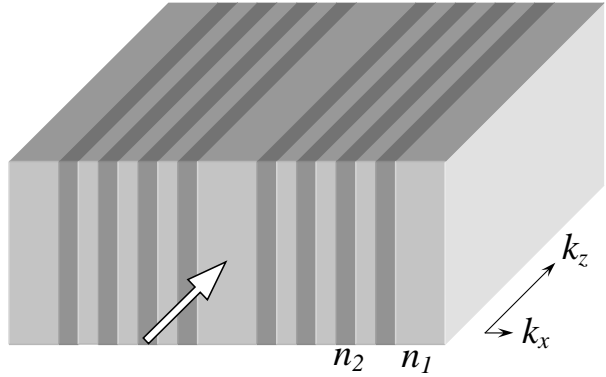}
\caption{A traditional ARROW waveguide. Refractive index
$n_2>n_1$. The big arrow denotes light is launched into the core
region. Notice the core can be made of a third material of
different index (normally lower than both $n_1$ and $n_2$)
\cite{Yeh:Multilayered}. $k_z$ is the wavevector component in
longitudinal direction, which can be more commonly written as
$\beta$. $k_x$ is the wavevector component in lateral direction.
Notice for a particular propagation mode, $k_x$ has different
values in layers with different refractive indices, whereas the
same $k_z$ value is shared in all layers.} \label{Fig:ARROW}
\end{figure}

Let's review what is the traditional ARROW model made popular by
Duguay \emph{et al.}. Refer to the simplest slab-type ARROW shown
in Fig. \ref{Fig:ARROW}, the model states that properties of the
high-index layers around the core determines the guidance in the
low-index core region. When the high-index cladding layers are in
resonance with the core mode, light is relayed outwards laterally
by the high-index layers. When the layers are not in resonance (in
antiresonance) with the core mode, light is rejected and
confinement in low-index core region is achieved. The resonant and
antiresonant conditions can be quantitatively derived from the
Bragg reflection conditions for the lateral wavevector component
$k_x$ (Fig. \ref{Fig:ARROW}). The lateral phase variation in a
high-index layer is
\begin{equation}
\phi_x=k_0\sqrt{n_2^2-n_1^2}\cdot d_2. \label{eq:phase}
\end{equation}
where $d_2$ is width of high-index layers. If $\phi_x$ is equal to
odd number of $\pi/2$, the high-index layers are considered to be
in antiresonant state with the core mode. Or we can say Bragg
reflection is met along $x$-direction. The core mode hence has the
least radiation loss under this condition. If $\phi_x$ is equal to
even number of $\pi/2$, the high-index layers are considered to be
in resonant state with the core mode. The core mode experiences
rapid dissipation after being launched.

This traditional ARROW model does explain successfully light
confinement in a material of lower index. It even does not
acknowledge the cladding to be periodic (which suggests that we
might have missed certain things as nowadays we explicitly use a
periodic PC cladding to confine light in a lower index material).
Also the model can accurately predict the exact highest-loss and
lowest-loss wavelength points by considering the high-index layers
in cladding to be in complete resonance and antiresonance,
respectively. However, we would like to point out that ARROW is
indeed just another name for regular waveguides whose guidance is
achieved by the photonic bandgap effect. Antiresonance of cladding
composite gives rise to a forbidden bandgap. And in fact, we
should establish a equation
\begin{eqnarray}
\mbox{cladding bandgap}&=&\mbox{completely antiresonant state of cladding}   \nonumber\\
&+&\mbox{partially antiresonant states of cladding}.
\label{eq:PBGARROW}
\end{eqnarray}

In the ARROW model proposed by Duguay \emph{et al.}, the role of
low-index layers in cladding region has been ignored. In fact,
their antiresonance also contributes to the confinement of
lightwave in core region\footnote{However, the antiresonance of
the low-index layers has less contribution than that of the
high-index layers to the confinement of lightwave. This can be
easily explained by calculating a few lowest-order supermodes (or
space-filling modes) of the cladding composite. The first group of
modes have about $\pi$ lateral phase variation in the high-index
layers. The next group of modes have about $\pi/2$ lateral phase
variation in the high-index layers. For both mode groups, very
little field is located in the low-index layers. If we plot the
dispersion curves for these two groups of modes in a $\beta$-$k$
diagram, the region in between of these two dispersion curve
groups is the bandgap used for our guidance. As little field is
located in low-index layers, this consequence is valid:
\begin{eqnarray}
&&\mbox{some small variation in the low-index layer width}   \nonumber \\
&\rightarrow&\mbox{little change in two mode groups' dispersion curves}  \nonumber \\
&\rightarrow&\mbox{little change in the bandgap}   \nonumber \\
&\rightarrow&\mbox{little change in guidance in core region}.
\nonumber
\end{eqnarray}
In above argument, we have assumed the primary bandgap of the
cladding is used for guidance. This makes sense, as ARROWs and
PBGFs fabricated to date almost explicitly use bandgaps found
among several lowest-order space-filling modes. }. In other words,
to better confine light in the core region, we need to tune the
widths of both high- and low-index layers in cladding so that they
meet their anti-resonance conditions together.

ARROW model is straightforward for slab-type waveguides shown in
Fig. \ref{Fig:ARROW}. Nice equations can be written down to
determine wavelength points where the lowest and the highest
leakage loss happen. However, it should be mentioned that such
equations are difficult, if not impossible, to be analytically
written down for a 2D PBG waveguide. For example, for the
all-solid PBG fiber shown in Fig. \ref{fig:MOFPBG}(f), though the
resonance conditions for the cladding high-index rods can be
approximately written down using certain expression
\cite{Litchinitser:ARROWOPEX,Yan:ICCCAS04}, resonance condition
for the low-index cladding region is not. Especially in such 2D
waveguides a full transverse bandgap of the cladding is necessary
for confining light. Hence we have to consider all wavevector
components in the transverse direction for analyzing resonance and
antiresonance conditions (remember we only considered
$x$-component of the wavevector for slab-type waveguide shown in
Fig. \ref{Fig:ARROW}).

At this point, we may conclude that the name ``photonic crystal
fiber'' is sometimes misleading, as it gives us the impression
that photonic bandgaps can only exist in periodic cladding
structures. A consequence of the ARROW model is that the cladding
of an PBG waveguide does not need to be periodic. Refer to Fig.
\ref{Fig:ARROW}, the most straightforward example would be: the
high-index cladding layers can be of certain width to cause
$\pi/2$ lateral phase variation, and they also can be thicker to
introduce $3\pi/2$ lateral phase variation. Besides, as far as the
optimum confinement at a particular wavelength is concerned, the
periodicity requirement of the cladding composite is automatically
lifted in the case of Bragg fiber (and any other 2D PBG
waveguide). This is owing to the fact that mode field in a Bragg
fiber is represented by aperiodic Bessel functions\footnote{For a
2D PBG fiber like Bragg fiber, we would however be better off by
using a periodic cladding in practice. There are several reasons.
First, aperiodic cladding can only enhance guidance at a single
wavelength point, not the whole cladding PBG window. And
improvement in PBG guidance can always be achieved by including
more cladding layers. Second, fabrication of periodic structure is
easier than that of aperiodic structure. Third, periodic structure
facilitates deployment of Bloch theorem in bandgap calculation.}
\cite{Yeh:BraggFiber}.

\section{PCF Classification}
\begin{figure}
\centering
\includegraphics[clip=true, width=12cm]{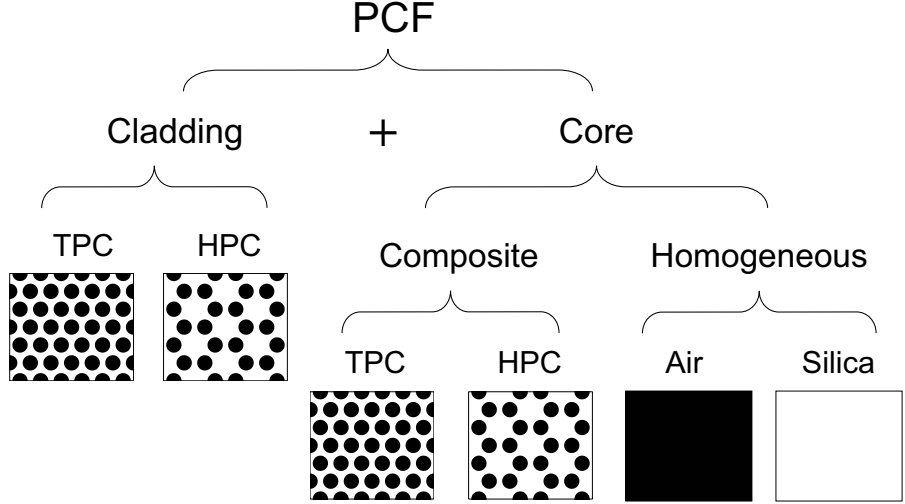}
\caption{Composition of a PCF. Black is for air, white is for
silica. TPC stands for ``triangular photonic crystal'' (air holes
are arranged in a triangular lattice). HPC stands for ``honeycomb
photonic crystal'' (air holes are arranged in a honeycomb
lattice).} \label{fig:classification}
\end{figure}

Figure \ref{fig:classification} shows the possible configurations
of a PCF. This figure has generalized PCF to those made of a PC
cladding and a PC core \cite{Yan:HSPCF}. Though we only consider
PCs made of two materials, it should be kept in mind that more
materials can be involved. Structurally, only PCs exhibiting
$C_{6v}$ symmetry are presented owing to their ease of
fabrication. Depending on fabrication method, other-latticed PCs
should be able to exist.

From Fig. \ref{fig:classification}, a particular two-material
fiber can be named accurately by mentioning the cladding and core
structures. For example, if the cladding is TPC and core is HPC,
then the fiber can be referred to as a TPC-HPC fiber; for a fiber
with TPC cladding and silica core, we can name it as solid-core
TPC fiber; for a fiber with TPC cladding and air core, we can name
it as air-core TPC fiber, etc.

Besides those shown in Fig. \ref{fig:classification}, we should
mention that a MOF can have a homogeneous cladding and a PC core
\cite{Yan:Antiguiding}.

\subsubsection{Triangular \emph{v.s.} Honeycomb}

These two terms are sometimes confusing. Early PCFs are made of
circular air holes in silica background. Hence, we normally tell a
PCF's cladding type according to the placement of air holes.
However, as PCF design gets complicated, such as air-guiding TPC
fiber and HPC fiber shown in Fig. \ref{fig:MOFPBG}(a) and (b), the
fiber cladding is more close to silica rods immersed in air
background \cite{Yan:HPCFAirCoreDesign,Yan:TPCImproved}. If we
define the lattice according to position of the silica pillars,
traditional
 TPC (HPC) becomes HPC (TPC).

Addressing the lattice pattern of a PC can also be ambiguous
without specifying corresponding lattice vectors. For a
traditional HPC, if we use a second-level unit-cell, the crystal
becomes a TPC \cite{Yan:HSPCF}.

In this thesis, we stick to traditional naming convention,
\emph{i.e.}, a crystal structure is defined by the air hole
positions, though the air holes may not be circular. Also, the
primitive basis vectors are normally used for defining the
crystal's lattice pattern, unless otherwise stated.

\section{Technology Review and Current Trends \label{Sec:Review}}

\subsection{Index-guiding MOF}

\subsubsection{Technology Review}

The possibility of fabricating a fiber whose cladding is made of
photonic crystal was first proposed in \cite{Birks:2DPC}
\footnote{Here we would like to draw attention to the paper
published in 1978 by Yeh \emph{et al.} on Bragg fiber
\cite{Yeh:BraggFiber}. Strictly speaking, their work should be
regarded as the first paper on \emph{fiber} design employing a
photonic crystal (1D) cladding. However, they did't generalize the
idea to 2D and 3D situations. Their work was put into practice by
a group of people in MIT about 20 years later (for example, see
\cite{Fink:Omniguide} and \cite{Temelkuran:BraggFiberNature}).
This is quite similar to the case that 1D photonic crystal was
first explained by Lord Rayleigh in 1887, but revived by E.
Yablonovitch \cite{Yablonovitch:PC} and S. John \cite{John:PC} 100
years later. There are always some people ahead of time.}. In this
paper, Birks \emph{et al.} theoretically showed that full 2D
photonic bandgaps exist in an air-silica (holes-in-silica)
photonic crystal (PC) when the off-plane wavevector component is
large enough. And they suggest ``a new type of optical fibre, the
PBG fibre, is possible'' by using ``a periodic air/silica cladding
surrounding a uniform core''. ``The core can be solid silica or a
hollow air region''. One year later, the same group published a
paper on the first fabricated photonic crystal fiber (PCF)
\cite{Knight:PCF96}. As the central defect is silica, the fiber
does not guide light using PBG effect. Though Knight \emph{et al.}
have included the term ``photonic crystal'' in naming the
waveguide, the fiber, in principle, is equivalent to a
conventional step-index fiber (SIF). Lack of PBG guidance,
however, does not prevent such fiber being exceptional. Knight
\emph{et al.} identifies an interesting property, \emph{i.e.}, it
has is a very broad single-mode operation wavelength range
(458nm$\sim$1550nm for the fiber reported, with $\Lambda=2.3\mu$m
and $d/\Lambda=0.15$). Their work attracted immediately lots of
attention and related studies were carried out by different
research groups around the world. More special waveguiding
properties were reported. In year 1998, Mogilevtsev \emph{et al.}
theoretically studied group velocity dispersion (GVD) of PCFs
\cite{Mogilevtsev:GVD}, and found some novel dispersion
properties, such as zero-dispersion point at short wavelength and
large normal dispersion value at 1550nm, can be easily achieved by
adjusting the fiber parameters ($\Lambda$ and $d$). The dispersion
tuning in such fibers was later overwhelmingly studied by
researchers. To name a few, dispersion-flatten PCFs were studied
in
\cite{Ferrando:DispersionFlattened,Ferrando:DispersionFlattenedOL,Reeves:DFPCF,Renversez:DFLowLoss,Saitoh:DispersionControlling},
dispersion-compensating PCFs were studied in
\cite{Birks:DCPCF,Shen:DCPCF,Zsigri:DCPCF,Gerorme:DCPCF}.
Remarkable birefringence property of PCFs were also reported on
several occasions in the literature. They include
\cite{Blanch:HIBIPCF,Hansen:HIBIPCF,Suzuki:PMPCF,Kubota:SinglePolarization}.
Nonlinear optics in PCFs is also demonstrated to be promising as
the fibers can be fabricated with ultrahigh nonlinearity can be
fabricated using large-sized cladding air holes and small silica
core. Among all nonlinear effects, third-harmonic generation in
PCF can be found in \cite{Efimov:THG}, four-wave mixing can be
found in, for example, \cite{Lee:FWM,Chow:FWM}, Raman effect can
be found in \cite{Yusoff:Raman,Matos:Raman,Fuochi:Raman},
supercontinuum phenomenon can be found in, for example,
\cite{Husakou:SC,Hori:SC,Hilligsoe:SC} etc, etc.

It should be mentioned that beside using silica, other materials
are also explored for, especially, index-guiding PCF design. The
materials include high-lead silicate glass, tellurite glass,
gallium lanthanum chalcosulfide glass, and also
polymethyl-methacrylate (PMMA). The first three types of glasses
have quite low softening temperature, hence they are also called
soft glasses. Using these soft glasses to fabricate PCFs has been
summarized in \cite{Feng:SGHF}. Two thermally-matched soft glasses
have also been used to form an index-guiding PCF \cite{Feng:SOHO}.
The first polymer PCF was reported in \cite{Eijkelenbour:PMOF}.

\subsubsection{Trends}
Though confinement loss of index-guiding MOF has been reduced to
0.48 dB/km \cite{Nielsen:LowestLoss}, we don't see immediate
necessity to deploy such fiber into transmission system. There are
several reasons for this: first of all, the loss is still higher
than conventional step-index fiber; second, the cost is certainly
higher as compared to standard commercial step-index
fiber\footnote{MOF uses only pure silica, therefore its preform is
cheaper in terms of material as compared to conventional SIFs
where doping of chemical element is involved. But stacking of
preform bundle is labor-intensive (there is currently no better
way to prepare the preform) and the uncertainties involved during
fabrication (dust contamination, fluctuation in pressure
controlling etc) is higher than that for conventional fiber.
Unless high-level automation is realized, it is correct to
conclude that MOFs are more expensive.}; third, compatibility
(such as splicing) with existing system is still a problem.
Index-guiding MOF can indeed be tailored to have some excellent
property in certain aspect, but not all aspects. In next few
years, application of such fibers would, most probably, still be
limited to subsystems or optical devices. In particular,
exploration of high nonlinearity and high birefringence properties
in MOFs would be likely to attract fair amount of attention. Its
deployment in fiber lasers will also be a very meaningful research
topic.

\subsection{PBG-guiding MOF}
\subsubsection{Technology Review}
The first PCF whose guidance is due to existence of its cladding
PC's photonic bandgap (PBG) was reported in year 1998 by Knight
\emph{et al.} \cite{Knight:PBGFScience}. However, the propagating
light concentrates in silica portion within the core region, which
is inevitable as the fiber's cladding is of honeycomb lattice and
the core is made of air-silica composite \cite{Yan:HSPCF}. The
first \emph{air-guiding} PBG fiber was reported in 1999 by Cregan
\emph{et al.} \cite{Cregan:AirguidingPBGF}. The transmission
spectrum is measured using a short fiber of a few centimeter
length. It is claimed to be single-mode. However, the authors
didn't convincingly prove their point. Fiber technique is greatly
improved in the following several years. By year 2003, loss is
greatly reduced to 13dB/km at 1500nm by Corning
\cite{Smith:TPCFCorning}. And in year 2004, lowest fiber loss is
even as small as 1.2 dB/km
\cite{Mangan:TPCFLargeCore,Robert:LowestLoss}.

Air-guiding PBG fiber was originally proposed for ultra-low loss
optical communication. In addition, it was said that signal
propagating in such fiber is ``free of group velocity
dispersion''. However, to date, we have achieved neither of these
two goals. The loss is still high compared to conventional
step-index fibers. The group velocity dispersion is quite large,
especially near bandgap edges \cite{Saitoh:TPCFLeakage}. Recently
it is found that there are uncertainties in such fiber's
polarization mode dispersion (PMD) property \cite{Wegmuller:PMD}.
What's more, a true low-loss single-mode air-guiding PBG fiber has
never been fabricated. All these factors would prevent such fiber
from being deployed for long-haul transmission link at this
moment. However, due to the unique hollow-core feature, it has
thrived in areas like high-power pulse delivery
\cite{Ouzounov:Magawatt,Shephard:HighPowerDelivery}, nonlinear
optics in gases \cite{Benabid:SBSScience,Benabid:Gascell}, atom
guiding \cite{Russell:PCF}, and sensing applications
\cite{Jenser:HollowCoreSensing}, etc.

Besides hollow-core PBG fibers, there exist several types of
solid-core PBG fibers. Theoretical work on this can be traced to
\cite{White:Resonance} and \cite{Yan:ICCCAS04}. Experimentally,
two soft glasses have been used to fabricate a PBG fiber by Luan
\emph{et al.} \cite{Luan:AllSolidPBGF}. Doped silica has also been
used to form scatterers in the cladding region of a solid-core PBG
fiber \cite{Aryros:OnePercent}. It was observed that such
low-index contrast PBG fiber has wider transmission window
\cite{Aryros:OnePercent}. However, no particular application has
so far been identified using these types of fibers, largely due to
the fact that light is confined in a solid core, which is quite
analogues to conventional step-index fiber.

Besides using 2D PC to form a PBG fiber's cladding, cylindrical-1D
PC was also exploited for confining light in a hollow core.
Omniguide fiber, or Bragg fiber, was first fabricated by Fink
\emph{et al} \cite{Fink:Omniguide}. The same group has
demonstrated using Bragg fiber to deliver high-power $\mbox{CO}_2$
laser \cite{Temelkuran:BraggFiberNature}. Recently, they have
produced such fiber with transmission window near IR (0.85 to
2.28$\mu$m) wavelength \cite{Kuriki:BFNIR}. Vienne \emph{et al.}
fabricated the so-called air-silica Bragg fiber whose cladding
highly resembles a cylindrical-1D PC \cite{Vienne:BFOPEX}. Very
similar Bragg fiber has also been fabricated very recently by
Agyros \emph{et al.} using air and polymer
\cite{Agyros:PrivateComm}. However, the air-silica Bragg fiber
suffers from a very fragmented transmission window
\cite{Vienne:BFOECC}. Whereas Bragg fibers made by Fink's group
has three materials (core is made of air, cladding is made of soft
glass and polymer bi-layers), people have also tried to fabricate
Bragg fibers using two solid materials (core material is one of
the cladding materials with lower index). These include
\cite{Brechet:PBGDSF,Katagiri:SilicaCorePBGF}.

\subsubsection{Trends}
The traditional air-guiding PBG fiber uses a TPC cladding.  Such
2D PC was recently generalized by Yan \emph{et al}
\cite{Yan:TPCImproved}. It is found that by slightly modifying the
PC structure, it is possible to tune the bandgap favorably for
different purposes, \emph{e.g.}, smaller leakage or even true
single-mode air-guiding etc. A hollow-core fiber with such PC
cladding was theoretically studied in \cite{Yan:ECOC05}. The fiber
has very wide surface-mode-free PBG guiding wavelength range and
it promotes single-mode operation. Technologically speaking,
fabrication of such fiber does not impose any additional
difficulty as compared to conventional air-guiding PBG fiber. It
will be very interesting to see an actual fiber in near future.

HPC cladding was not considered to achieve air guiding until its
capability was theoretically validated by Yan \emph{et al.}
\cite{Yan:HPCFAirCore}. Such fiber's leakage loss was later
studied in \cite{Yan:HPCFAirCoreDesign}. The hollow-core of such
fiber can be less than 6$\mu$m in diameter, which suggests
single-mode operation is possible. Also the surface-mode problem
has been completely mitigated in such fibers. Air-guiding HPCF can
be a competitive alternative to existing air-guiding PBG fibers if
they can be fabricated successfully.

As we have mentioned, the first fabricated honeycomb PBG fiber has
an air-silica composite core \cite{Knight:PBGFScience}. The idea
of using two different pieces of PCs to form a fiber (core and
cladding are both composite) was put forward by Yan \emph{et al.}
\cite{Yan:HSPCF}. Depends on the design, PBG-guiding or
index-guiding can be achieved in such fibers. Such fibers are
expected to perform better in application of gas or liquid sensor.
Also, they should have important applications that exploit the
phenomenon of discrete soliton \cite{Neshev:VortexSolitons}.

As compared to air-silica PBGFs, Bragg fiber suffers significantly
higher radiation loss. New types of high index-contrast,
low-absorption dielectric materials which are matched in thermal
properties (softening temperatures, expansion coefficients,
viscosity) and chemical properties (non-diffusive to each other)
need to be identified to fabricate more robust hollow-core Bragg
fiber. Preferably, the guidance should be around 1550nm wavelength
point. For air-silica Bragg fiber fabricated by Vienne \emph{et
al.}, more theoretically work is necessary to understand its
leakage loss. It might be possible to prevent the fragmentation of
the transmission window by an improved design. If that indeed
happens, we can reduce the overall leakage loss by increasing the
number of air-hole rings in the cladding.

Application-wise, high-power pulse or beam delivery will still be
an important direction to go \cite{Limpert:MOFHighPower}.
Nonlinear optics in such fibers with gas- or liquid-filled core
will also be very interesting.

\bibliographystyle{osajnl}
\bibliography{PCFbib}

\end{document}